\documentclass[aps,prl,twocolumn,superscriptaddress,showpacs,floatfix]{revtex4-1}

\usepackage{color}
\usepackage{graphicx}
\usepackage{amsmath}
\usepackage{amssymb}
\usepackage{bm}
\usepackage{array}
\newcolumntype {s}[1]{@{\hspace{#1}}} 
\usepackage{colordvi}

\graphicspath{{.}{./EPS/}}

\newcommand* {\ee}{\ensuremath{\mathrm{e}}}

\newcommand*{\vek}[1]{{\ensuremath{\bm{\mathrm{#1}}}}}

\begin{document}

\title{Soliton Magnetization Dynamics in Spin-Orbit Coupled Bose-Einstein Condensates}

\author{O. Fialko}
\affiliation{Centre for Theoretical Chemistry and Physics and New Zealand Institute for
Advanced Study, Massey University, Private Bag 102904 NSMC, Auckland 0745, New
Zealand}

\author{J. Brand}
\affiliation{Centre for Theoretical Chemistry and Physics and New Zealand Institute for
Advanced Study, Massey University, Private Bag 102904 NSMC, Auckland 0745, New
Zealand}

\author{U. Z\"ulicke}
\affiliation{School of Chemical and Physical Sciences and MacDiarmid Institute for Advanced 
Materials and Nanotechnology, Victoria University of Wellington, PO Box 600, Wellington
6140, New Zealand}

\begin{abstract}
Ring-trapped Bose-Einstein condensates subject to spin-orbit coupling support localized dark soliton excitations that show periodic density dynamics in real space.  In addition to the density feature, solitons also carry a localized pseudo-spin magnetization that exhibits a rich and tunable dynamics. Analytic results for Rashba-type spin-orbit coupling and spin-invariant interactions predict a conserved magnitude and precessional motion for the soliton magnetization that allows for the simulation of spin-related geometric phases recently seen in electronic transport measurements.


\end{abstract}


\maketitle

The recent realization of artificial light-induced gauge potentials for neutral atoms~\cite{spiel:nat:09}
has added a powerful new instrument to the atomic-physics simulation toolkit~\cite{dalib:rmp:11}. In
particular, possibilities to induce Zeeman-like and spin-orbit-type couplings in (pseudo-)spinor atom
gases~\cite{spiel:nat:11} render them ideal laboratories to investigate the intriguing interplay of spin
dynamics and quantum confinement that has been the hallmark of semiconductor
spintronics~\cite{lossbook,zutic:rmp:04}. At the same time, the unique aspects of
Bose-Einstein-condensed atom gases~\cite{pitaev03} associated, e.g., with their intrinsically
nonlinear dynamics, promise to give rise to novel behavior under the influence of synthetic
spin-orbit couplings~\cite{stan:pra:08,merkl:prl:10,wang:prl:10,ho:prl:11,wu:cpl:11,yip:pra:11,sinha:prl:11,hu:prl:12,zhang:prl:12}.

One of the special properties resulting from nonlinearity in Bose-Einstein condensates
(BECs) is the existence of solitary-wave excitations~\cite{gonz:nonl:08}. Basic types of
these are distinguished by the shape of their localized density feature: dark (gray) solitons
are associated with a full (partial) depletion of a uniform condensate density in a finite region
of space, whereas bright solitons are localized density waves on an empty background.
A further characteristic associated with solitons is the phase gradient of the condensate
order parameter centered at the position of the density feature. In multi-component systems,
the dynamics of soliton excitations is found to be enriched by the additional degrees of
freedom~\cite{mana:jetp:74,kivshar:pre:97,ohb:prl:01,smy:pra:10}.

\begin{figure}[b]
\vspace{-0.5cm}
\includegraphics[height=5cm]{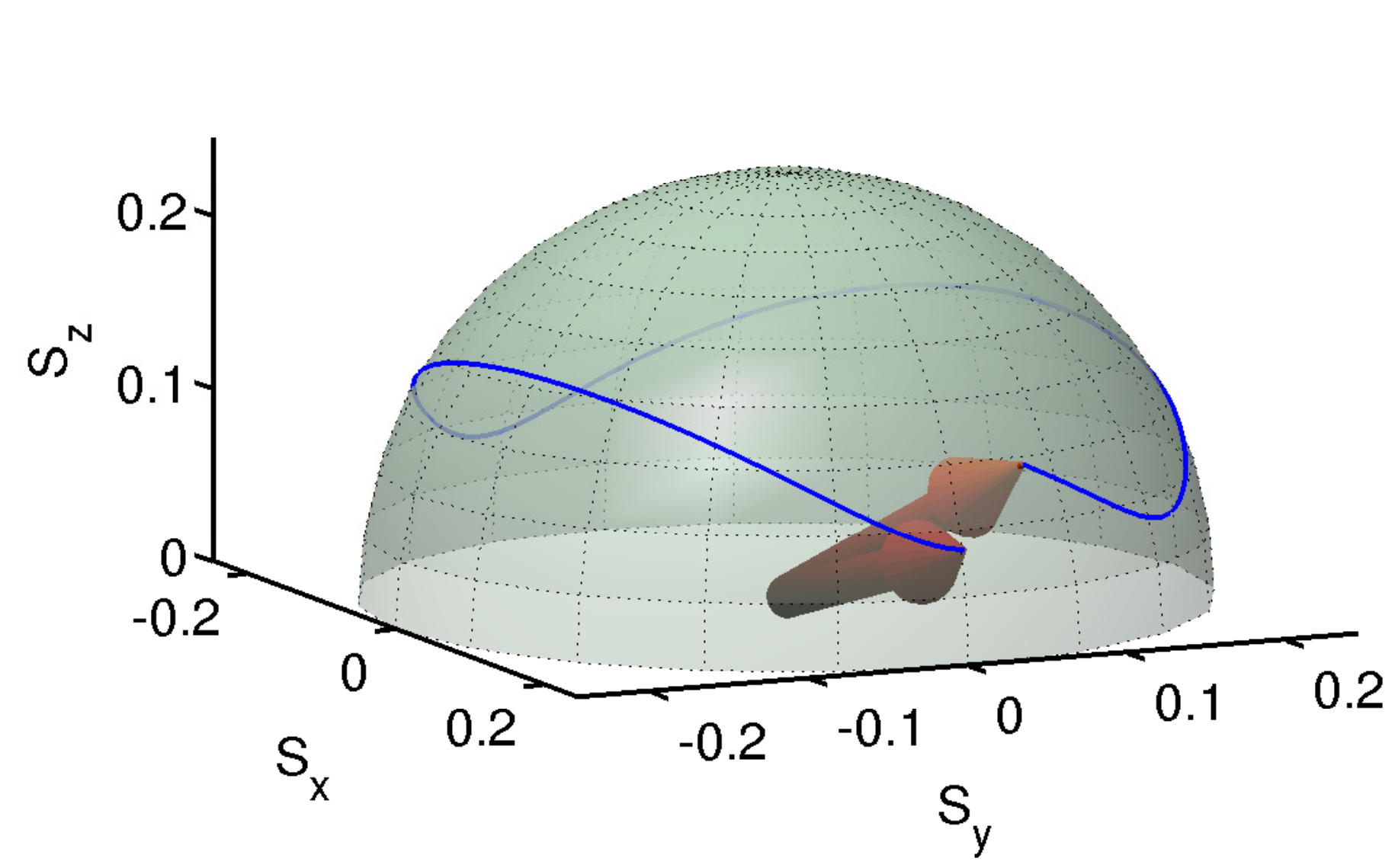}
\caption{\label{fig:GBdyn1}
Time evolution of a gray-bright soliton's magnetization in a ring-trapped BEC with Rashba
spin-orbit coupling. During a full cycle of the soliton's motion on the ring, the magnetization
vector follows a trajectory on the surface of a sphere. The magnetization vectors at
the beginning and the end of a cycle (indicated by arrows) differ by an angle
$\vartheta$ that is related to a spin-related geometric phase. Soliton parameters (see text):
$v_{\text{s}}/c=0.5$, $\tan\eta=2$, $g=100$, $\kappa=-0.01$.}
\end{figure}

We have studied solitons in ring-trapped pseudo-spin-$1/2$ condensates 
with spin-invariant repulsive atom-atom interactions
subject to a Rashba-type~\cite{rashba:60,byra:jpc:84} spin-orbit coupling and find that they exhibit
a third feature: a pseudo-magnetization vector with conserved magnitude and rich dynamics
that unfolds in tandem with the soliton's periodic propagation in real space. Figure~\ref{fig:GBdyn1}
shows an example and also illustrates the interesting fact that the magnetization directions at
the beginning and the end of a full cycle of the soliton's motion are generally not parallel. The
appearance of such a geometric phase~\cite{wilczekBook} and the precessional time evolution
of the solitonic magnetization is reminiscent of the spin dynamics of electrons traversing a
mesoscopic semiconductor ring~\cite{loss90,yuli:prl:93,uz:prb:03b,frust:prb:04,nitta:prl:12}.

In the following, we consider several soliton configurations and obtain analytical results for their
density and magnetization profiles as well as the magnetization dynamics associated with their
motion. We start by introducing the basic theoretical description of our system of interest. Using
the basis of a spatially varying \emph{local spin frame\/}~\cite{uz:prb:03b} for the condensate
spinor, the nonlinear Gross-Pitaevskii equation~\cite{pitaev03} for the spin-orbit-coupled
ring BEC turns out to be of Manakov-type~\cite{mana:jetp:74}, making it possible to apply
standard methods~\cite{kivshar:pre:97,smy:pra:10} to find solitary-wave solutions. Accounting for
the presence of spin-orbit coupling adds an important twist: Spinors have to satisfy non-standard
boundary conditions, which introduce background-density flows in the local spin frame that
contribute to the nontrivial magnetization dynamics exhibited by the moving solitons in the lab frame.

We consider a two-component BEC trapped in the $xy$ plane and confined to a ring of radius $R$.
The atoms are assumed to be in the lowest quasi-onedimensional subband~\footnote{This
limitation is not crucial, as finite-width effects and higher subbands could be treated
straightforwardly.} and subject to a spin-orbit coupling of the familiar Rashba
form~\cite{rashba:60,byra:jpc:84} $\alpha_{\text{R}} \left[\sigma_x (-i\partial_y) - \sigma_y (-i
\partial_x)\right]$ as well as a spin-rotationally invariant contact interaction. (Here $\sigma_{x,y,z}$
are the spin-1/2 Pauli matrices.) The energy functional of such a system~\cite{merkl:epjd:10} is
given by $E[\Psi]=\int d\varphi \,\,\Psi^{\dag} \,({\mathcal H} -\mu)\,\Psi$, where 
$\varphi$ is the azimuthal angle, $\mu$ the
chemical potential, $\Psi =(\psi_\uparrow, \psi_\downarrow)^T$ the two-component
(pseudo-spin-$1/2$) spinor order parameter in the representation where the ($z$)
direction perpendicular to the ring's plane is the spin-quantization axis, and 
\begin{eqnarray}
\nonumber
{\mathcal H} &=& E_0 \bigg[ -\partial_\varphi^2 +\frac{g}{2}\,\Psi^\dag \Psi  \\
 && \hspace{0.5cm} + \tan\eta \left( \sigma_+ \ee^{-i\varphi} + \sigma_- 
 \ee^{i\varphi}\right)\left(-i \partial_\varphi +\frac{\sigma_z}{2}\right)\bigg] .
\end{eqnarray}
We use $\sigma_\pm\equiv (\sigma_x\pm i\sigma_y)/2$ to denote raising and lowering
operators for spin-1/2 components, $E_0 = \hbar^2/(2 M R^2)$ is the energy scale
for  quantum confinement of atoms with mass $M$ in a ring of radius $R$,
$E_0 g$ is the two-body contact-interaction strength, and $\tan\eta = 2 M R\,
\alpha_{\text{R}} /\hbar^2$ is a dimensionless measure of the spin-orbit coupling.

The effect of Rashba spin-orbit coupling in a ring geometry can be elucidated by
performing a suitable SU(2) transformation. Defining $\Psi =
{\mathcal U}\,\chi$ and ${\mathcal H}_{\text{loc}} = {\mathcal U}^{-1}
\mathcal H\,{\mathcal U}$, with ${\mathcal U}= \ee^{- i\varphi
\sigma_z/2}\,\ee^{i {\eta} \sigma_y/2}\, \ee^{i\varphi\sigma_z/(2\cos\eta)}$, we find
\begin{equation}\label{eq:LocRG}
{\mathcal H}_{\text{loc}} = E_0 \left[ -\partial_\varphi^2  - \frac{(\tan\eta)^2}{4}
+ \frac{g}{2}\,\chi^\dagger\chi \right] \quad .
\end{equation}
The transformation ${\mathcal U}^{-1}$ amounts to a $\varphi$-dependent
rotation of the pseudo-spin quantization axis~\cite{uz:prb:03b}, followed
by a spin-dependent gauge transformation. We will refer to the original
representation where the spin-quantization axis coincides with the axis of the ring as
the \textit{lab frame\/}, whereas the representation in which the Hamiltonian of the
system is diagonal in pseudo-spin space [i.e., given by ${\mathcal H}_{\text{loc}}$ of
Eq.~(\ref{eq:LocRG})] will be the \textit{local spin frame\/}~\cite{uz:prb:03b}.
Note that the spinors $\Psi$ in the lab frame are periodic functions of $\varphi$,
whereas the spinors $\chi=(\chi_+, \chi_-)^T$ from the local spin frame have to satisfy
the boundary conditions $\chi_\pm(\varphi)=\chi_\pm(\varphi+2\pi)e^{\pm i {\mathcal A}}$
with a spin dependent phase twist originating from the spin-orbit coupling, where
\begin{equation}\label{eq:AAphase}
{\mathcal A} = \pi\left( \frac{1}{\cos\eta} -1 \right) \quad .
\end{equation}

Knowledge of the local-spin-frame spinors enables the calculation of expectation values
for any observables accessible to measurement in the lab frame. The total density $n =
|\psi_\uparrow|^2+|\psi_\downarrow|^2 \equiv |\chi_+|^2+|\chi_-|^2$ is obviously the same
irrespective of which representation is chosen in spin space. The pseudo-spin-1/2
projections in the lab frame correspond to definite atomic states, hence their density
profiles $n_{\uparrow(\downarrow)} = \Psi^\dagger ([1+ (-) \sigma_z]/2)\Psi$ are of
interest. In addition, we will consider  the magnetization-density vector $\vek{s} =
\Psi^{\dag}\vek{\sigma}\, \Psi$ in the lab frame, with $\vek{\sigma}=(\sigma_x,\sigma_y,
\sigma_z)$ being the vector of Pauli matrices. 

We analyze the properties of localized excitations in spin-orbit-coupled
ring-trapped BEC based on the time-dependent Gross-Pitaevskii equation~\cite{pitaev03}
$\delta E[\chi]/\delta\chi_{\sigma}^\ast = i\hbar \,\partial\chi_{\sigma}/\partial t$. After rescaling to use the
dimensionless time variable $\tau = t E_0 /\hbar$, it has the form
\begin{equation}\label{eq:nonlGP}
i\, \frac{\partial \chi_\sigma}{\partial \tau} = \left[-\partial^2_\varphi + g\,\left(\left|\chi_+
\right|^2 + \left|\chi_-\right|^2 -n_0\right)\right] \chi_\sigma 
\end{equation}
for the two components of the spinor $\chi = (\chi_+, \chi_-)^T$, where $n_0 = [\mu+
(\tan\eta)^2/4]/(g E_0)$ is the uniform (background) density consistent with
the chemical potential $\mu$. While the spin-orbit coupling has formally disappeared
from the nonlinear equation (\ref{eq:nonlGP}), it is still implicitly present via the boundary
conditions that the individual components $\chi_\pm(\varphi,\tau)$ must satisfy.

\begin{figure*}[t]
\includegraphics[height=4cm]{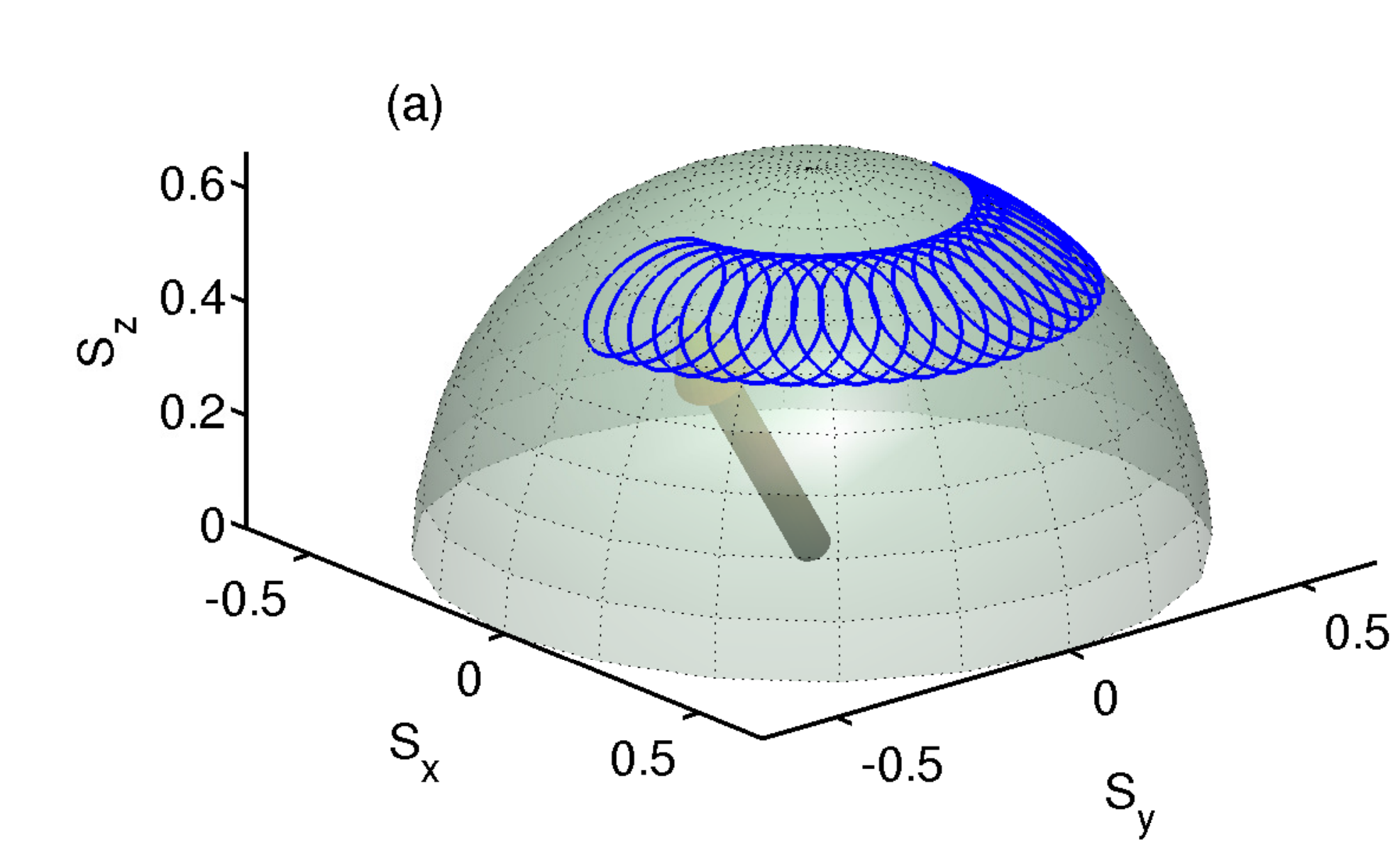}
\includegraphics[height=4cm]{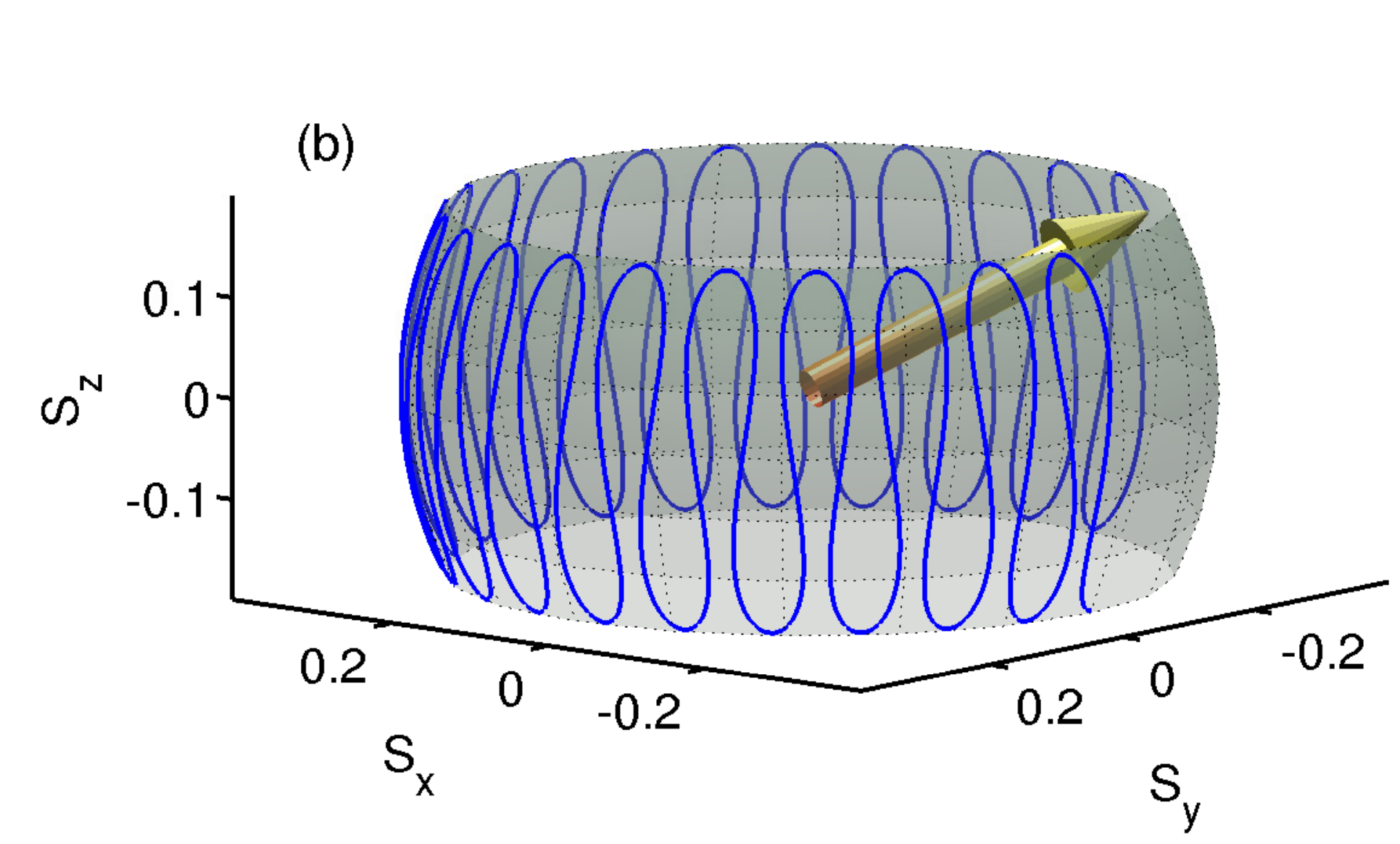}
\includegraphics[height=4cm]{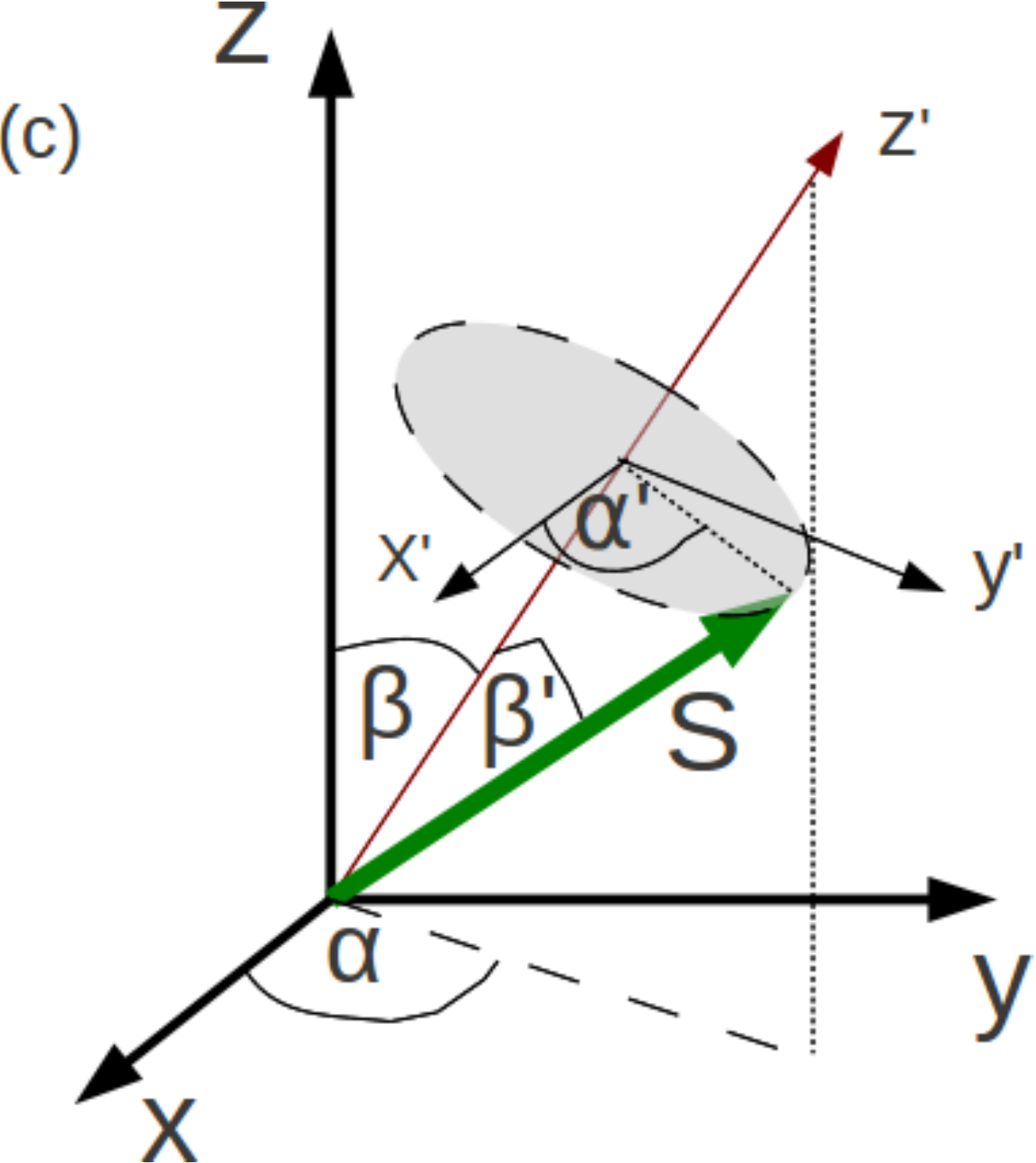}
\caption{\label{fig:SolDyn}
Magnetization dynamics of a gray-bright soliton [panel (a)] and a gray-gray soliton with zero
background magnetization in the local spin frame [panel (b)]. Parameters used are $g=100$
($100$), $\tan\eta=0.2$ ($0.5$), $v_{\text{s}}/c=0.5$ (0.2), and $\kappa=-0.5$ for the gray-bright
(gray-gray) case. (c)~Angles used to describe the two-step precessional motion of $\vek{S}$. The angle $\alpha'$ is measured with 
respect to an $x'$ axis that is perpendicular to both the $z$ and $z'$ axes.}
\end{figure*}

We have obtained several soliton solutions of Eqs.~(\ref{eq:nonlGP}) using established
techniques~\cite{mana:jetp:74, kivshar:pre:97,smy:pra:10} and implemented the
appropriate boundary conditions. Before giving further details, we like to summarize a few
general features. The soliton spinors in the local-spin-frame representation turn out to be
of the form
\begin{equation}\label{eq:genSol}
\chi_\sigma^{(\text{s})} = \Upsilon^{(\text{s})}_\sigma(\varphi-v_{\text{s}}\tau)\,\ee^{i
v_{\text{b}\sigma} {\varphi}/2-i v_{\text{b}\sigma}^2\tau/4} \quad,
\end{equation}
where $\Upsilon^{(\text{s})}_\sigma(\xi)$ are complex amplitude functions encoding the
specific soliton-like density features, $v_{\text{s}}$ is the propagation speed of the soliton,
and $v_{\text{b}\sigma}$ are background flow velocities of the individual spinor components
that are necessary to implement the boundary conditions arising due to the presence of
spin-orbit coupling. 
The density $n_{\uparrow(\downarrow)}^{\text{(s)}}$ and magnetization density $\vek{s}^{\text{(s)}}$
exhibit spatially localized features. Subtracting
$\vek{s}^{\text{(s)}}$ from the magnetization density $\vek{s}_{\text{b}}^{\text{(s)}}$ of the
condensate background yields the magnetization density that is associated with the soliton
excitation only. Its integral $\vek{S}^{\text{(s)}} = \int d\varphi\,\, [\vek{s}_{\text{b}}^{\text{(s)}} -
\vek{s}^{\text{(s)}}]$ is the vector of total soliton magnetization, which is an additional property
of localized excitations in multi-component BECs. For soliton solutions of the form (\ref{eq:genSol}), $\vek{S}^{\text{(s)}}$ has constant magnitude. 
Its temporal evolution is most conveniently described by a set of four angles
as defined in Fig.~\ref{fig:SolDyn}(c). 
While the tilt angles $\beta$ and $\beta'$ are time-independent, the angles $\alpha$ and $\alpha'$ vary linearly in time, 
signifying the precession of $\vek{S}^{\text{(s)}}$ around tilted $z'$ axis with the universal result
\begin{equation}\label{eq:univAngles}
\beta=\eta, \quad \alpha = v_{\text{s}}\tau + \pi .
\end{equation}
The $z'$ axis is tilted by the angle $\eta$ characterizing the spin-orbit coupling and it rotates around the $z$ axis with the same
angular velocity $v_{\text{s}}$ that characterizes the soliton propagation.
The second tilt angle $\beta'$ is found to depend only on the soliton profile $\Upsilon_{\pm}(\varphi)$, while the precession frequency
$d\alpha'/d\tau$ has complicated dependences on the parameters of the soliton solutions. 
Figure~\ref{fig:SolDyn} shows exemplary magnetization dynamics for
gray-bright and gray-gray solitons. Interestingly, we find that the magnetization
vector is usually not parallel to its initial direction after the soliton has completed a full cycle of its
motion around the ring as, e.g. seen in figure \ref{fig:GBdyn1}. The angle $\vartheta$ between the magnetization
directions at the start and the end of a cycle turns out to be finite only as a consequence of spin-orbit coupling,
as it depends prominently on the phase ${\mathcal A}$ given in Eq.~(\ref{eq:AAphase}) that also
governs spin-dependent interference in mesoscopic ring conductors~\cite{frust:prb:04}.

In order to find explicit soliton solutions, we introduce $\xi={\varphi}-u\tau$, where $u$ is a velocity
parameter, and initially look for solutions of the form $\chi_\sigma(\varphi, \tau)=\sqrt{n_\sigma(\xi)}
\, \ee^{i\theta_\sigma(\xi)}$. This allows us to rewrite Eq.~(\ref{eq:nonlGP}) in the form
\begin{subequations}\label{eq:solitonDE}
\begin{eqnarray}
\label{Eq.cont}
-u\frac{\partial n_\sigma}{\partial\xi}+2\frac{\partial}{\partial\xi}\left(n_\sigma\frac{\partial
\theta_\sigma}{\partial\xi}\right)=0\, , \qquad \\ \label{Eq.euler} 
u\frac{\partial \theta_\sigma}{\partial\xi}+\frac{1}{\sqrt{n_\sigma}}\frac{\partial^2
\sqrt{n_\sigma}}{\partial^2\xi}-\left(\frac{\partial\theta_\sigma}{\partial\xi} \right)^2 - g (n-n_0)=0 
\, , \qquad
\end{eqnarray}
\end{subequations}
where $n=n_++n_-$. Single component solutions for $\sigma=\tilde{\sigma}$ are easily found by
integration of Eqs.~(\ref{eq:solitonDE}) to yield $\chi_{\tilde{\sigma}}(\xi) \propto\Upsilon(\xi)$ and
$\chi_{-\tilde{\sigma}}=0$, with the well-known dark soliton solution on the infinite line~\cite{pitaev03}
\begin{equation}\label{Eq.fulln1}
{\Upsilon(\xi)=\sqrt{n_{0}}\, \left[i\frac{u}{c}+\gamma_u \tanh\left(\gamma_u\,\frac{\xi}
{\xi_{\text{D}}}\right)\right]\quad .}
\end{equation}
Here $\gamma_u^2=1-u^2/c^2$, $c^2=2 g n_{0}$, $1/\xi_{\text{D}}^2= g n_{0}/2$. The soliton profile
(\ref{Eq.fulln1}) is appropriate for sufficiently strong nonlinearity, where $\xi_D/\gamma_u\ll2\pi$
\footnote{Otherwise periodic solutions involving elliptic functions have to be used. See, e.g.,
L.~D. Carr, C.~W. Clark, and W.~P. Reinhardt, Phys. Rev. A \textbf{62}, 063610 (2000).}. However,
$\Upsilon(\xi)$ does not satisfy the proper boundary condition since it has a phase step $\Delta
\theta =-2\arccos(u/c)$. To compensate for the phase  step and ensure the correct phase shift
associated with the gauge transformation ${\mathcal U}$, we perform a Galilean transformation on
(\ref{Eq.fulln1}), which yields
\begin{equation}
\chi^{\text{SC}}_{\tilde{\sigma}}({\varphi},\tau)=\Upsilon(\xi-v_{\text{b}}\tau)\,
\ee^{i v_{\text{b}}{\varphi}/2-i v_{\text{b}}^2\tau/4}\quad . 
\label{Eq.Galileyn1}
\end{equation}
Here $v_{\text{b}}=-(\Delta\theta + {\tilde{\sigma}}{\cal A})/\pi$ is the background velocity imposed
by the boundary condition. Thus the single-component soliton solution is of the form
(\ref{eq:genSol}), with {$\Upsilon^{\text{SC}}_{\tilde{\sigma}} = \Upsilon$ and
$\Upsilon^{\text{SC}}_{-\tilde{\sigma}} = 0$, and propagation speed $v_{\text{s}}=u+v_{\text{b}}$. }

A straightforward calculation yields $\vek{s}^{\text{SC}} = \tilde{\sigma} | \Upsilon(\varphi -
v_{\text{s}} \tau)|^2 (-\sin\eta\,\cos\varphi, -\sin\eta\,\sin\varphi,\cos\eta)^T$ for the
magnetization density of the single-component soliton solution. In essence, the density
depletion at the soliton's position gives rise to a reduction of the magnetization density
$\vek{s}_{\text{b}}^{\text{SC}} = \tilde{\sigma}\, n_{0} (-\sin\eta\,\cos\varphi,-\sin\eta\,\sin
\varphi,\cos\eta)^T$ associated with the background. Thus $\vek{s}_{\text{b}}^{\text{SC}} -
\vek{s}^{\text{SC}}$ constitutes the magnetization density associated with the soliton itself, as it
is the change in the background magnetization density due to the presence of the localized
excitation. For the single-component soliton, this corresponds to a peak in magnetization
density at the soliton's position. The total magnetization vector is obtained by integrating that
peak in real space, which yields $\vek{S}^{\text{SC}} = \tilde{\sigma} (-\sin\eta\,\cos(v_{\text{s}}\tau),
-\sin\eta\,\sin(v_{\text{s}}\tau),\cos\eta)^T$. This magnetization vector is precessing in a perfectly
synchronized fashion with the soliton's motion around the ring [cf.\ Fig.~\ref{fig:SolDyn}(c)
with Eq.~(\ref{eq:univAngles}) and $\beta'=0$], i.e., $\vartheta^{\text{SC}}=0$.

\begin{figure}[t]
\hspace{0.15cm}\includegraphics[width=.46\columnwidth]{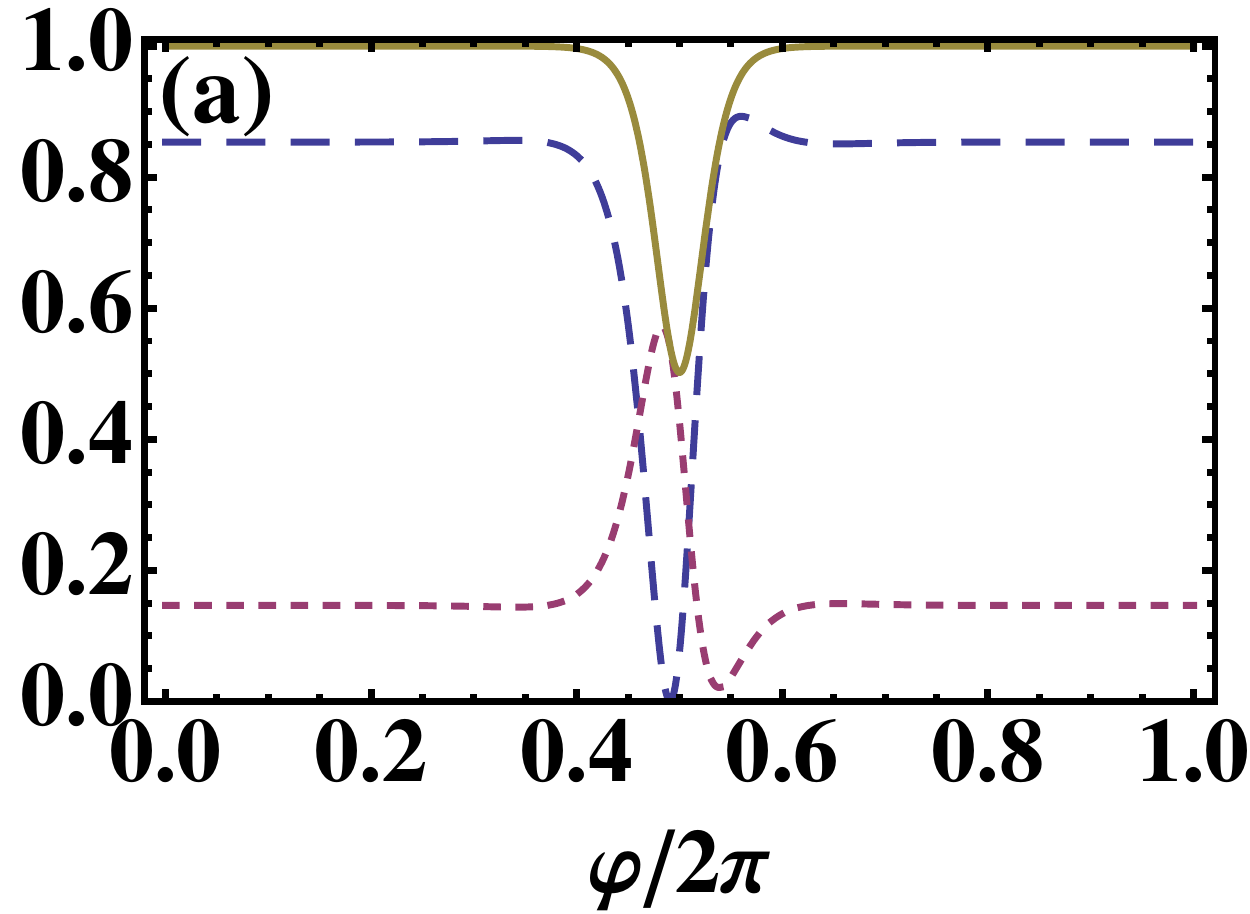}\hfill
\includegraphics[width=.46\columnwidth]{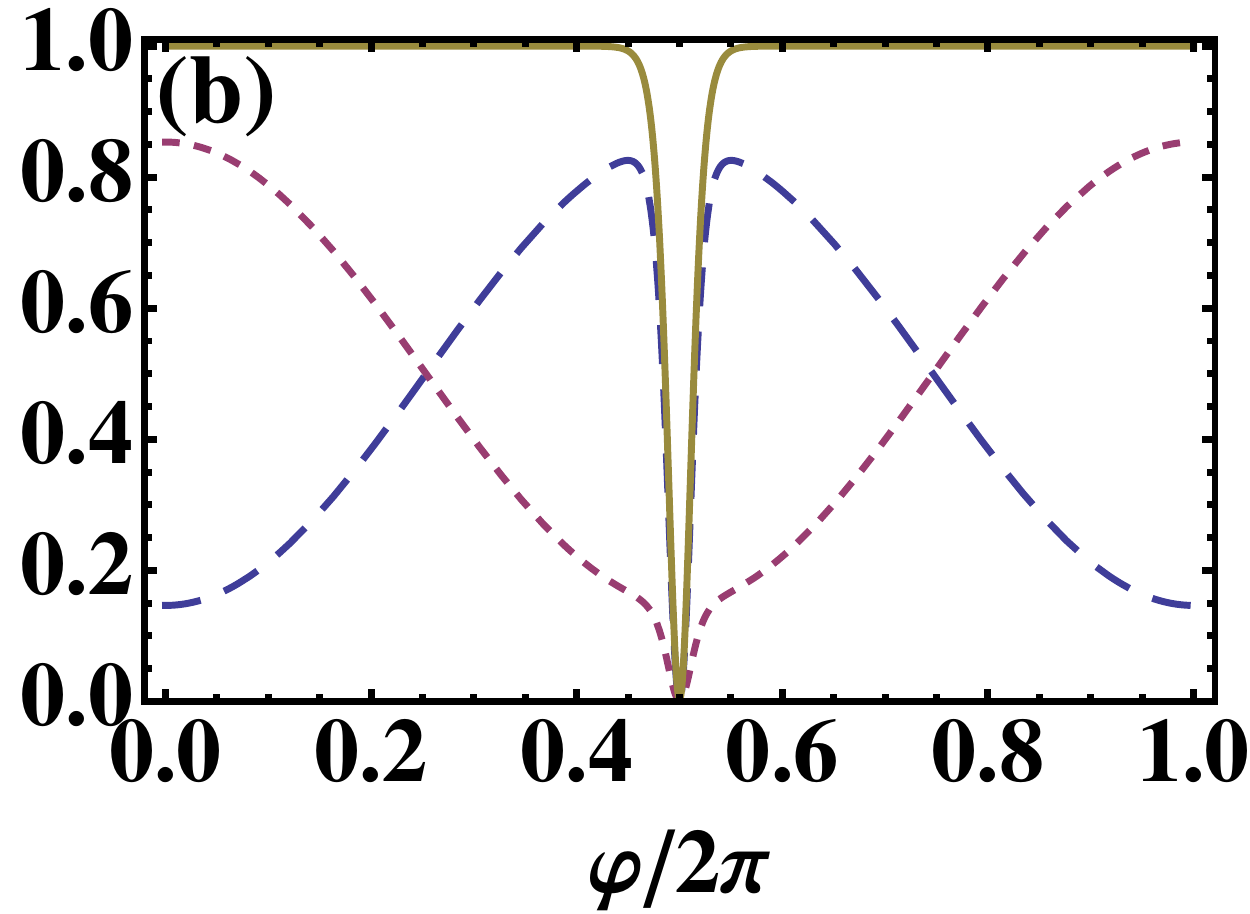}\\
\includegraphics[width=.48\columnwidth]{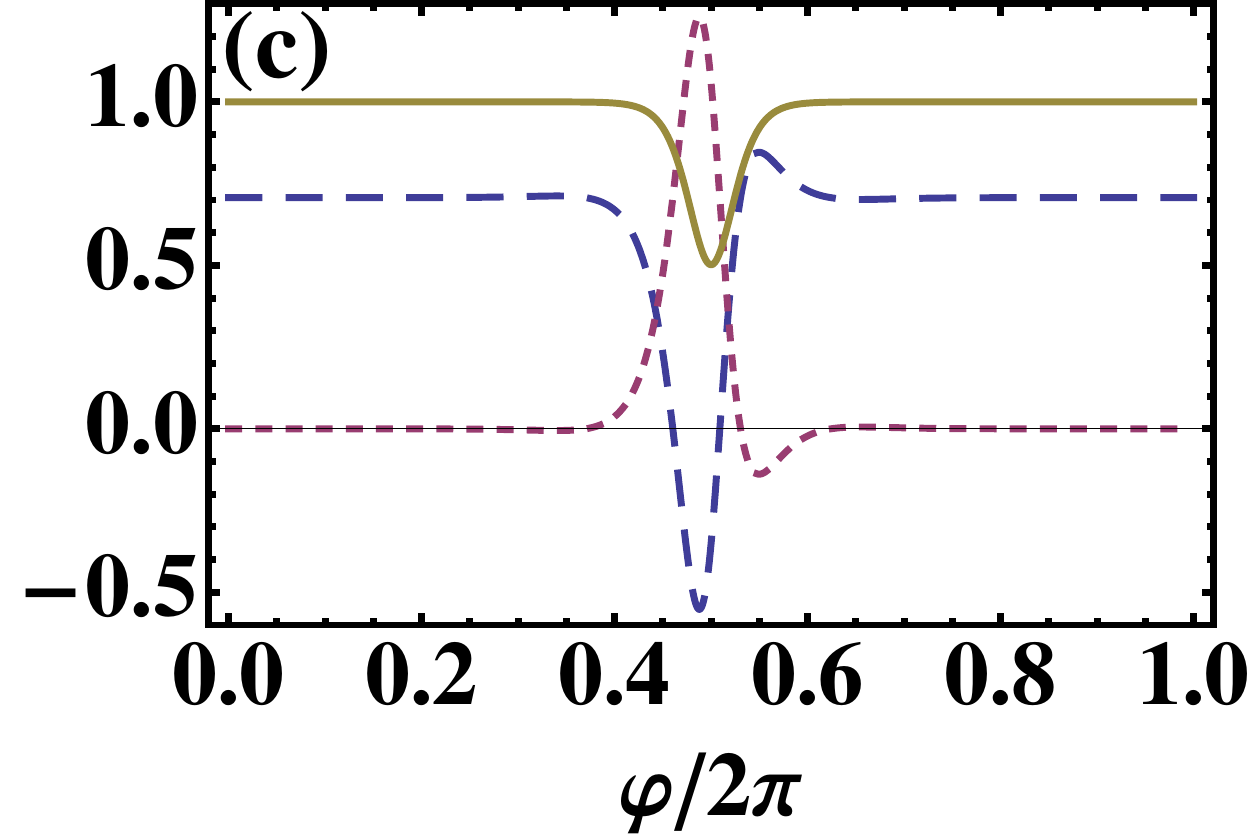}\hfill
\includegraphics[width=.48\columnwidth]{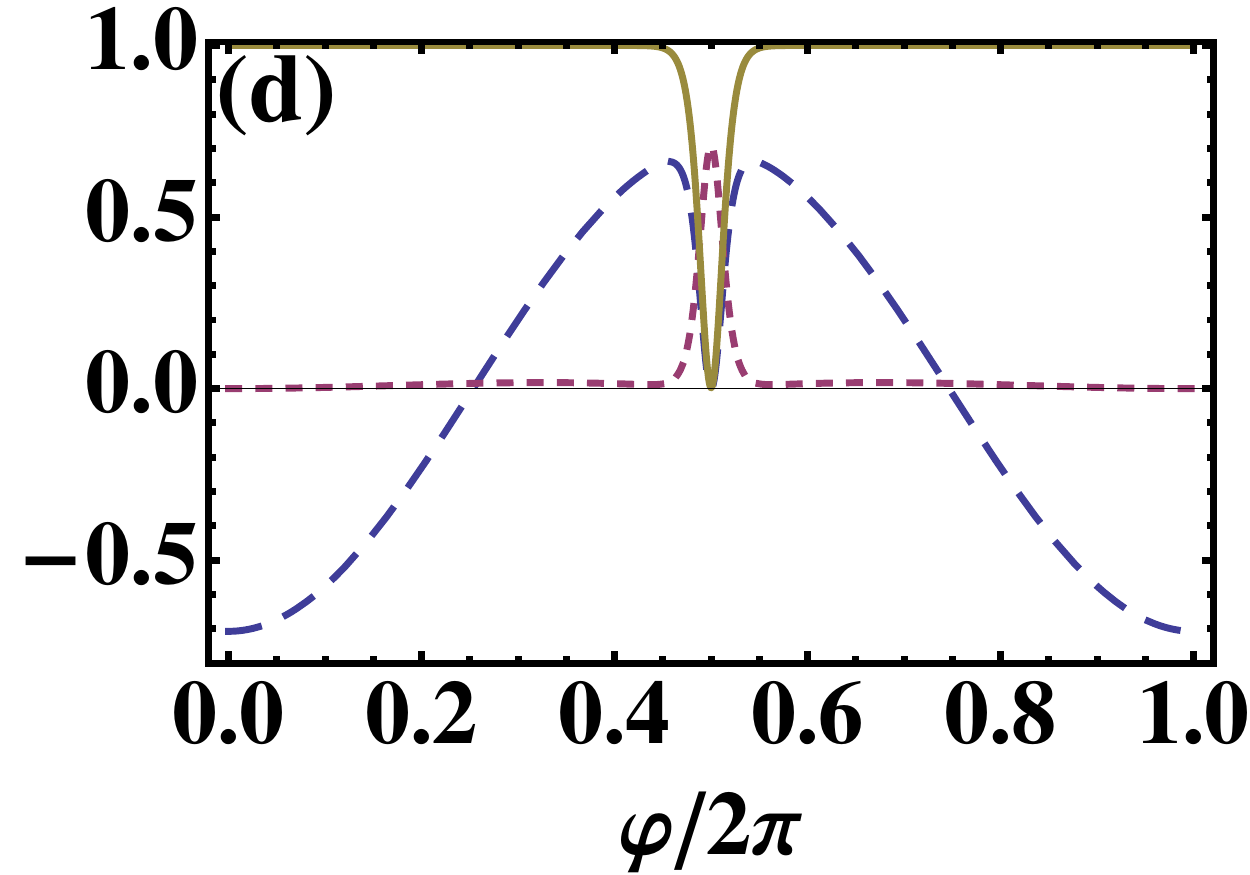}
\caption{\label{fig:profiles}
Lab-frame spin densities [(a),(b)] and $z$ component of the magnetization density [(c),(d)] for
a stationary gray-bright soliton [(a),(c)] and a stationary gray-gray soliton [(b),(d)]. Panels (a)
and (b) show the total density (solid yellow curve) and individual-spin (dashed blue $=$ $\uparrow$,
dotted red $=$ $\downarrow$) densities normalized to the background-density value. In (c)
and (d), the total density is plotted again for reference as the solid yellow curve, together with the
magnetization profiles $s_z^{\text{(s)}}$ (dashed blue curve), and the magnetization density
$s^{\text{(s)}}_{\text{b}\, z} - s^{\text{(s)}}_z$ associated with the soliton excitations only (dotted
red curve). Parameters are $\tan\eta=1.0$, $g=100$, and (for the gray-bright soliton) $\kappa=-0.5$.}
\end{figure}

We now consider a solution of Eqs.~(\ref{eq:solitonDE}) that is a gray-bright (GB) soliton in the
local spin frame. We assume that the densities approach constant values $n_+\rightarrow n_{0+}$
(gray part) and $n_- \rightarrow 0$ (bright part) far away from the soliton's position. To decouple
Eq.~(\ref{Eq.euler}), we use the ansatz~\cite{smy:pra:10} $n_-=\kappa (n_+-n_{0+})$ with $-1\le 
\kappa \le 0$. We apply a Galilean boost to both components to match the phase of the gray part
only, hence they are of the form (\ref{eq:genSol}) with $\Upsilon^{\text{GB}}_+(\xi)$ given by
$\Upsilon(\xi)$ from Eq.~(\ref{Eq.fulln1}) but with rescaled $c^2=2gn_{0+} (1+\kappa)$,
$1/\xi_{\text{D}}^2=g n_{0+}(1+\kappa)/2$, and
\begin{equation}
\Upsilon^{\text{GB}}_-(\xi)= \sqrt{-\kappa n_{0+}}\, \frac{\gamma_u\, \ee^{i u\xi/2}}{\cosh
(\gamma_u \xi /\xi_{\text{D}})} \quad .
\end{equation}
Furthermore, $v_{\text{b}-}=v_{\text{b}+}$ and $v_{\text{s}} = u + v_{\text{b}+}$.
Figure~\ref{fig:profiles} shows the density profiles [panel (a)] and magnetization-density profile
[panel (c)] associated with a GB soliton.

The vector $\vek{S}^{\text{GB}}$ of total magnetization for a GB soliton precesses concomitantly
with the soliton's motion; cf.\ Fig.~\ref{fig:SolDyn}(c) with Eq.~(\ref{eq:univAngles}) and
$\tan\beta'=(\sqrt{-\kappa}u\pi)/[(1-\kappa)c \gamma_u]$, $\alpha'= -v_{\text{s}}\tau(1+{\cal A}/\pi)$.
Figures~\ref{fig:GBdyn1}(a) and \ref{fig:SolDyn}(a) show examples of possible time evolutions of
the GB-soliton magnetization. The magnitude of the magnetization vector is found to be $S = 2(1-
\kappa)\xi_D\gamma_u/\cos\beta'$. For a GB soliton, the magnetization vector turns out to be
\emph{not\/} aligned with its initial direction after completion of a full cycle of its motion around the
ring. A straightforward calculation yields $\sin(\vartheta^{\text{GB}}/2) = \sin\beta'\, \sin{\mathcal A}$.
As $\beta'$ is a known function of the soliton parameters, a measurement of $\vartheta^{\text{GB}}$
will yield the spin-related geometric phase ${\mathcal A}$.

The solutions representing gray-gray (GG) solitons in the local spin frame are obtained by Hirota's
method~\cite{kivshar:pre:97}. The spinor components are of the form (\ref{eq:genSol}) with
\begin{equation}
\Upsilon_\sigma^{\text{GG}}(\xi)=\sqrt{n_{0\sigma}}\left[i\frac{u_\sigma}{c_\sigma}+
\gamma_{u_\sigma}\tanh(a\xi) \right] \quad .
\label{Eq.ChiGG}
\end{equation}
Here, $a^2=\gamma_{u_+}^2/\xi_{\text{D}+}^2+\gamma_{u_-}^2/\xi_{\text{D}-}^2$,
$v_{\text{s}}=2a u_\sigma/c_\sigma\gamma_{u_\sigma} + v_{\text{b}\sigma}$, $c_\sigma^2 
=2 g n_{0\sigma}$, $1/\xi^2_{\text{D}\sigma}= g n_{0\sigma}/2$. The back-ground
flows are given by $v_{\text{b}\sigma}=-(\Delta\theta_\sigma +  \sigma {\cal A})/\pi$, where $\Delta\theta_\sigma = - 2\arccos(u_\sigma/c_\sigma)$. 
The independent parameters characterizing a GG soliton are the ratio $n_{0+}/n_{0-}$ (or,
equivalently, the background magnetization in the local spin frame) and the speed
$v_{\text{s}}$ of the soliton. All other parameters can be found by solving transcendental
equations given just after Eq.~(\ref{Eq.ChiGG}). For simplicity, we consider the case of a
GG soliton with zero background magnetization in the local spin frame (i.e., $n_{0+}=n_{0-}
\equiv n_0/2$). Figure~\ref{fig:profiles} shows results for spinor-density [panel (b)] and
magnetization-density [panel (d)] profiles.

The time evolution of the magnetization vector associated with a moving GG soliton is
characterized by the angles defined in Fig.~\ref{fig:SolDyn}(c) with
Eq.~(\ref{eq:univAngles}), $\beta' = \pi/2$, and $\alpha'=\omega\tau + \pi/2$, where
$\omega=-(v_{\text{b}+}-v_{\text{b}-}) v_{\text{s}}/2+(v_{\text{b}+}^2-v_{\text{b}-}^2)/4-
(1+{\cal A}/\pi)v_{\text{s}}$. Figure~\ref{fig:SolDyn}(b) illustrates this dynamics which,
for small $\eta$, corresponds to a slow rotation of the magnetization vector in the ring's
plane with superimposed fast small-amplitude oscillations in the normal direction. As in
the case of the GB soliton, the magnetization vector does not evolve back to its initial
direction after a period of the soliton's ring revolution. The angle between magnetizations
at the start and the end of the cycle is found to be $\vartheta^{\text{GG}} = 2\pi |\omega|/
v_{\text{s}}\rightarrow 2{\mathcal A}$ for ${\mathcal A}\ll 1$. Again, the dependence of
$\vartheta^{\text{GG}}$ on ${\mathcal A}$ enables determination of the latter by measuring
the former.

In conclusion, we have investigated the properties of soliton excitations in ring-trapped
spin-orbit-coupled BECs. We find that a magnetization degree of freedom is generally
associated with a soliton, and that the magnetization vector precesses around an axis
that is rotating synchronously with the soliton's orbital motion around the ring. The
magnetization direction at the end of a cycle of revolution does not coincide with the initial
direction for the gray-bright and gray-gray cases, making it possible to measure a
spin-orbit-related geometric phase. Our work opens up new avenues for the realization and
manipulation of  magnetic soliton excitations in BECs. It also creates the opportunity to study
spin-dependent interference and scattering effects that, until now, were only accessible in
semiconductor nanostructures.

This work was supported by the Marsden fund (contract no.\ MAU0910) administered by the Royal Society of New Zealand.

%

\end{document}